\documentclass[conference]{IEEEtran}

\usepackage{amssymb}
\usepackage{amsmath}
\usepackage{balance}

\usepackage[colorlinks=true, linkcolor=blue, citecolor=blue, urlcolor=blue]{hyperref}

\usepackage{graphicx}
\usepackage{booktabs}
\usepackage{multirow}
\usepackage{tabularx}
\usepackage{threeparttable}
\usepackage{url}

\begin{document}

\title{Empirical Notes on the Interaction Between Continuous Kernel Fuzzing and Development}
\author{
\IEEEauthorblockN{Jukka Ruohonen}
\IEEEauthorblockA{University of Turku, Finland \\ 
Email: juanruo@utu.fi}
\and
\IEEEauthorblockN{Kalle Rindell}
\IEEEauthorblockA{SINTEF Digital, Norway \\
Email: kalle.rindell@sintef.no}
}

\maketitle

\begin{abstract}
Fuzzing has been studied and applied ever since the 1990s. Automated and continuous fuzzing has recently been applied also to open source software projects, including the Linux and BSD kernels. This paper concentrates on the practical aspects of continuous kernel fuzzing in four open source kernels. According to the results, there are over 800 unresolved crashes reported for the four kernels by the \textit{syzkaller/syzbot} framework. Many of these have been reported relatively long ago. Interestingly, fuzzing-induced bugs have been resolved in the BSD kernels more rapidly. Furthermore, assertions and debug checks, use-after-frees, and general protection faults account for the majority of bug types in the Linux kernel. About 23\% of the fixed bugs in the Linux kernel have either went through code review or additional testing. Finally, only code churn provides a weak statistical signal for explaining the associated bug fixing times in the Linux kernel.
\end{abstract}

\begin{IEEEkeywords}
Linux, BSD, crash, software vulnerability
\end{IEEEkeywords}

\section{Introduction}

Fuzzing is essentially a software testing technique that supplies invalid or random input to a software system with the goal of crashing the system. It has long been an important technique for improving software security~(see \cite{ChenCui18} for a comprehensive review of the history and associated literature). Recently, fuzzing of operating system kernels has received considerable attention~\text{\cite{Carabas17, Corina17, HanCha17, Schumilo17}}. Another recent trend has been the adoption of frameworks for continuous and automated fuzzing of various open source projects. Despite of these advances, only little research has been done for better understanding the practical software engineering aspects; fuzzing is not only about finding crashes, but also about debugging, fixing, and triaging bugs that may have serious security implications. These practical aspects provide the motivation for the present work---as well as its contribution. In fact, the paper is presumably the first to examine the software engineering aspects of the recently introduced continuous kernel fuzzing frameworks for the Linux and BSD~operating system kernels.

While a ``major goal of fuzzing and any bug discovery effort is to find as many bugs as possible''~\cite{ZhaoLiu16}, this goal is only a part of the larger picture. The bugs found should be also understandable and fixable with a reasonable amount of effort. Given time and resource constraints, then, the benefits from fuzzing should be evaluated also against the ``costs related to finding and fixing security related bugs''~\cite{Takanen09}. There are a couple of important words embedded to the previous quotation: costs and security. In commercial software development the costs involved often refer to concrete money. Although ``the cost per defect is always less than the cost of a security compromise'', such monetary aspects imply different business calculations, including considerations about the return on investment and the total cost of ownership~\cite{Takanen08}. Indirect costs are present also in open source software development. If there is a shortage of human resources and time---as is typically the case in most open source software projects, prioritization is often necessary in case a fuzzer outputs a large amount of crashes that may be difficult and time-consuming to understand and debug. Security is the obvious criterion for a prioritization; crashes with a clear security impact should have a priority. However, it is far from being straightforward to determine whether a fuzzing-induced crash has explicit security implications.

Analogous cost-benefit-security issues have recently sparked a lively debate in the open source community~\cite{LWN19a}. The debate has also touched the question about whether Common Vulnerabilities and Exposures (CVEs) should be mass-filed for the crashes found by the automated frameworks that are nowadays continuously fuzzing many open source projects. The CVE allocation question implicitly carries also another tenet: the Linux kernel development culture has a long history in explicitly or implicitly hiding security issues in commit messages~\text{\cite{LWN08a}}. Historically many security bugs in the Linux kernel have also lacked CVE identifiers~\cite{Wijayasekara12}, which has made triaging and prioritization even more burdensome for many stakeholders, including Linux~distributions~\cite{syzbot18b}. 

Even though CVEs are not the topic of the present work, the practical focus is still maintained: the paper's goal is to explore bug fixing times, bug types, and related software engineering aspects in the context of continuous kernel fuzzing. To this end, three research questions are examined about crashes and bugs triggered by \textit{syzkaller}---a hybrid kernel fuzzing framework developed by Google and associates~\cite{syzkaller19a}. The dataset is based on the corresponding \textit{syzbot} online dashboard for tracking the crashes triggered by \textit{syzkaller}~\cite{syzbot19a}. Although the primary focus is on the Linux kernel, also three BSD kernels are briefly examined with respect to bug fixing~times. 

The structure of the paper's remainder is straightforward. Namely: the opening Section~\ref{sec: research design} outlines the research design, including the research questions examined and a brief motivation for these; Section~\ref{sec: results} presents the empirical results; and the final Section~\ref{sec: discussion} concludes with a few remarks about the answers reached, limitations, and directions for further work.

\clearpage
\pagebreak
\section{Research Design}\label{sec: research design}

\subsection{Research Questions}

Three research questions are examined. The first is:

\begin{itemize}
\item{RQ.1: \textit{(a)~How long does it take to fix kernel bugs triggered by kernel fuzzing, (b)~and do the bug fixing times vary across Linux, FreeBSD, NetBSD, and OpenBSD in terms of empirical cumulative distribution functions?}}
\end{itemize}

This question is easy to justify. Bug fixing times are a classical topic in software engineering. Although not all bugs found by fuzzing are security bugs, previous results generally indicate that also security bugs often take a surprisingly long time to fix in many different contexts~\cite{Ruohonen19EASE, Ruohonen18IST}. Some bugs are never fixed even though these have been recognized as vulnerabilities~\cite{Gorbenko17}. No universal explanation is known for these and related results. Numerous different explanations are offered in the literature: some build on bug triaging aspects and  different incentives for vendors, bug reporters, and developers~\text{\cite{Ruohonen19EASE, Hosseini12}}; others stem from bug severity, testing, architectural flaws, dependencies, code complexity, and code churn~\text{\cite{Chinthanet19, Lamkanfi12}}; some are related to problems in vulnerability disclosure and associated coordination, including the allocation of CVEs for the vulnerabilities~\text{\cite{Ruohonen18IST, Muegge18}}; and so forth. Whatever the explanations may be, the first research question is worth asking to better understand the time delays associated with continuous fuzzing and automated testing in general. As for the corollary question RQ.1b, the four kernels mentioned again constitute a classical setup in empirical software engineering~\cite{Spinellis08}. Given the different history of the four kernels and their different development cultures, a basic hypothesis is that the three BSD kernels differ from the Linux kernel also in terms of fixing times for fuzzing-induced bugs. 

The second research question examined adds more context:

\begin{itemize}
\item{RQ.2: \textit{(a)~What types of bugs have been found in the Linux kernel through kernel fuzzing, (b)~and how often the already fixed bugs have been reviewed and tested?}}
\end{itemize}

The types of crashes triggered are generally interesting. Many of these are specific to the C programming language and kernel development. These are relevant also for deducing about the potential security implications. By using the Common Weakness Enumeration (CWE) framework as a reference, the examples include such security-related bug classes as  NULL pointer dereference \text{(CWE-476)}, double frees \text{(CWE-415)}, uninitialized variables (CWE-457), use-after-free (CWE-416), out-of-bounds reads (CWE-125) and writes (CWE-787), time-of-check-time-of-use (CWE-367), and various different deadlock and race conditions (CWE-366, CWE-667, and CWE-833, among others). Many of these bug classes have also been typical to the Linux kernel throughout the decades~\cite{Palix14}. Given that code reviews are extensively used in the Linux kernel development~\cite{Rigby14}, it can be expected that also crashes reported by \textit{syzbot} are at least sometimes reviewed and tested by developers other than a given committer or a subsystem maintainer. Thus, both RQ.2a and RQ.2b are worth asking.

The third research question launches a small probe into the potential explanations behind the bug fixing times in Linux:

\begin{itemize}
\item{RQ.3: \textit{Do the average bug fixing times in the Linux kernel vary (a)~according to bug types (RQ.2a), (b)~according to whether these have been reviewed and tested (RQ.2b), and (c) do these further correlate with code churn?}}
\end{itemize}

This question is again easy to motivate with the help of the noted literature on (security) bug fixing times. For instance, different memory management bugs (such as CWE-415 and CWE-416) could be assumed to yield faster fixing times compared to stalls and deadlocks that are usually difficult to debug especially in the kernel development context. In general, a similar rationale applies to the coordination of CVE identifiers and vulnerabilities in general~\cite{Ruohonen18IST}. Of course, testing and reviewing take time, and thus RQ.3b implies a clear-cut hypothesis. Finally, code churn in RQ.3c refers to commit-by-commit churn (such as files modified or lines of code added) that inevitably occurs when fixing  bugs. The Linux kernel is also famous for the extensive amount of code churn~\cite{Palix14}. Furthermore, different churn metrics have been used for predicting bug-prone classes, files, or commits, including cases where the bugs have been identified as vulnerabilities~\text{\cite{Meneely13, Nagappan05, ShinMeneely11}}. This background justifies also RQ.3c.

\subsection{Data}

The dataset is based on a snapshot collected in 30th of June 2019 from the \textit{syzbot} online dashboard~\cite{syzbot19a}. In general, \textit{syzbot} refers to the software engineering or automation layer for the \textit{syzkaller} fuzzer. Together with compiler instrumentation, this template-based and coverage-guided fuzzer relies on different kernel sanitizers (see \cite{syzbot18a} and \cite{LWN16b} for gentle introductions to the technical details). The notable kernel sanitizers include (but are not limited to) the KASAN (KernelAddressSANitizer), KMSAN (KernelMemorySanitizer), and KTSAN (KernelThreadSanitizer) implementations. While the technical details are interesting, the paper's practical software engineering focus warrants three additional remarks about the data used.

First, the three BSD kernels are examined only with respect to RQ.1. The reason is simple: due to different version control and bug tracking systems, it is unclear how well the BSD systems have so far been integrated into the \textit{syzbot} infrastructure. (Revealingly, a few fixed bugs in the BSD kernels also miss links to the corresponding commits in the online dashboard.) The same point applies to the reporting functionality embedded to the infrastructure. For the Linux kernel \textit{syzbot} tries to also automatically report the crashes to the respective subsystem maintainers~\cite{syzbot18a}, but a similar functionality seems absent for the BSD kernels. A further point can be made with respect to RQ.2b; the code review practices differ between the Linux and the BSD kernels~\cite{Rigby14}, and, thus, direct comparability is problematic. Second, RQ.3b and RQ.3c are examined only with respect to already fixed bugs in the Linux kernel. The reason is again practical: while \textit{syzbot} tries to bisect the crashes to the specific commits that introduced the bugs behind the crashes, it remains unclear how well the bisection functionality works in practice. Therefore, the already fixed bugs provide a more robust data source because manual (human) evaluation has been present. Third, there are one-to-many relations between commits and fuzzing-produced crashes; a single commit may fix multiple bugs. To deal with the issue, the evaluation of RQ.2b and RQ.3 operates at the bug-level. On one hand, this choice is justifiable because it avoids aggregation that a commit-level analysis would necessitate; on the other hand, some of the commit-based values are replicated across multiple observations.

\subsection{Methods}

The first two research questions are examined with descriptive statistics. The classical Kolmogorov-Smirnov test is used for evaluating the question RQ.1b. The identification of bug types for RQ.2 and RQ.3 is based on simple string matching that is easy to apply thanks to the more or less structured format outputted by the (Linux) kernel sanitizers~\cite{syzbot18a}. Regarding RQ.2b and RQ.3b, the evaluation is based simply on searching the \texttt{reviewed-by}, \texttt{tested-by}, and \texttt{reviewed-and-tested-by} character strings from the lower-cased commit messages, as outputted by \texttt{git}'s \texttt{format=\%B} command line option. As for RQ.3c, it is worth remarking that operationalization of code churn metrics varies from a study to another~\cite{Meneely13, Nagappan05}. Therefore, it seems also justified to use \texttt{git}'s \texttt{shortstat} option for the commit logs.

Two classical regression methods are used for RQ.3. First, the average (mean) bug fixing times across the bug types identified are checked with analysis-of-variance (ANOVA). The corollary questions entail a choice about a suitable regression estimation strategy. In addition to ordinary least squares, the typical choices in the domain include Poisson regression and its variants, quantile regression, and survival analysis~\cite{Ruohonen18IST}. The survival approach is used in the present work by using the Cox's classical proportional hazards regression model~\cite{Cox72}, which, however, is closely related to the Poisson regression in applied research~\cite{Callas98}. In terms of computation, the \textit{car} package is used for the ANOVA checks~\cite{FoxWeisberg11}, and the standard R functions \texttt{coxph} and \texttt{cox.zph} for the Cox's regression.

\section{Results}\label{sec: results}

\subsection{Bug Fixing Times (RQ.1)}

\begin{figure}[th!b]
\centering
\includegraphics[width=8cm, height=4.7cm]{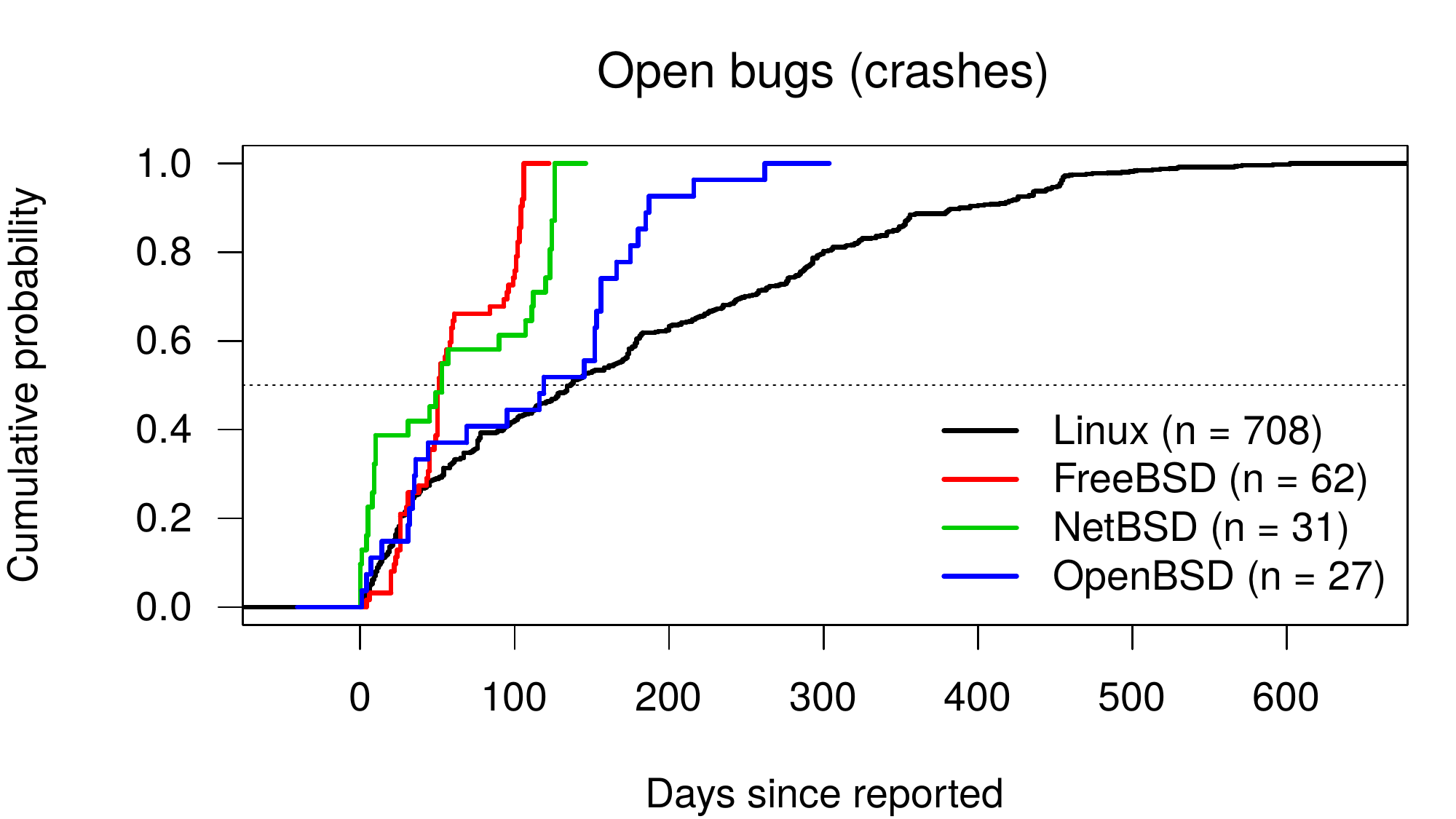}
\caption{Pending Bugs (Crashes) Across Four Kernels}
\label{fig: reported}
%
\vspace{15pt}
%
\centering
\includegraphics[width=8cm, height=4.7cm]{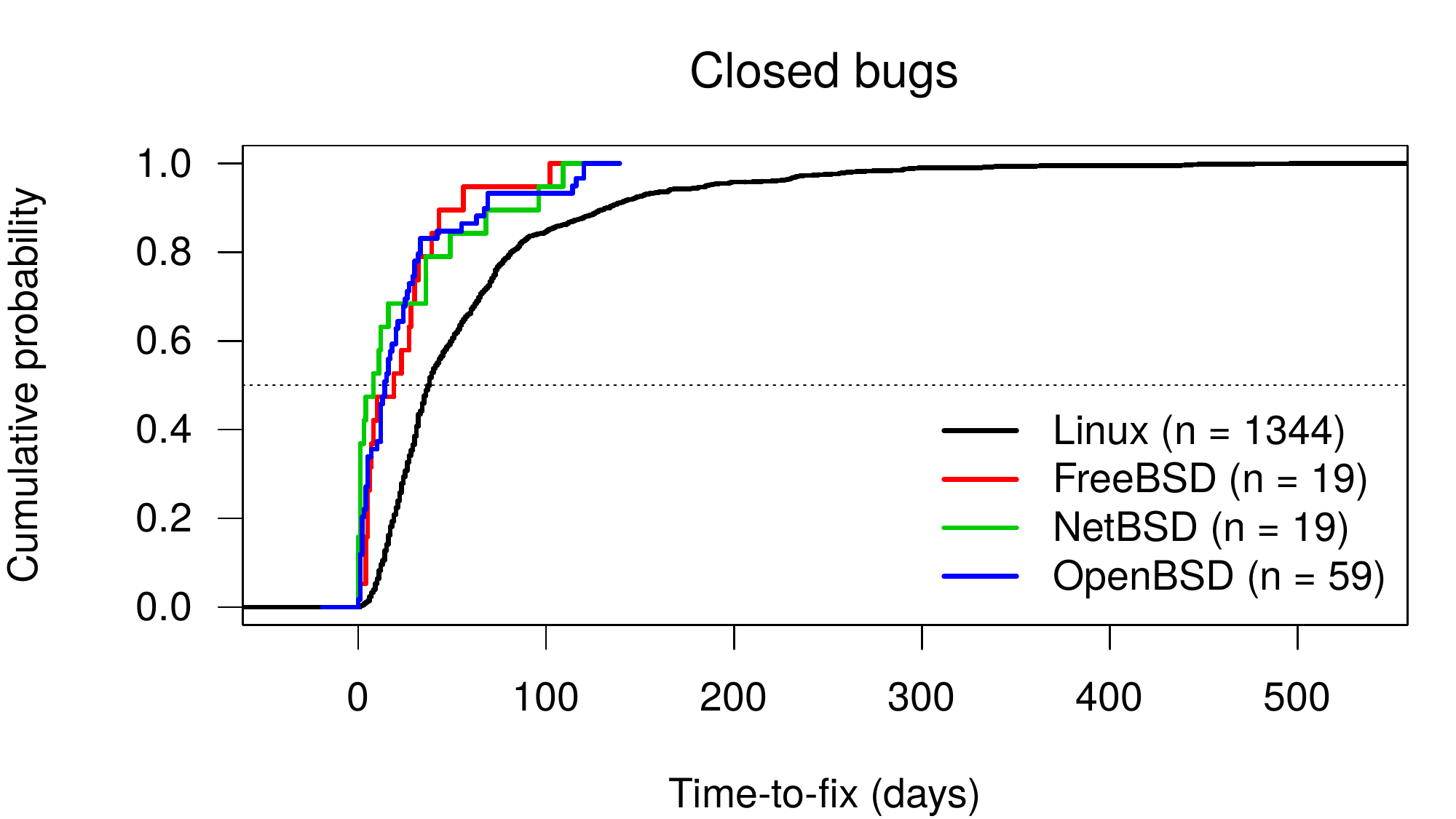}
\caption{Bug Fixing Delays Across Four Kernels}
\label{fig: ttf}
\end{figure}

The question about bug fixing time delays can be approached with two metrics: (a)~``\textit{days-since-reported}'' (DSR), that is, the days passed after \textit{syzbot} first reported the still open crashes, and (b)~``\text{\textit{time-to-fix}}'' (TTF) difference available by subtracting the days passed since the closed bugs were fixed from the DSR values. This operationalization is common in the domain~\cite{Muegge18}. The corresponding empirical cumulative distribution functions are shown in Fig.~\ref{fig: reported} and Fig.~\ref{fig: ttf}, respectively. 

\begin{table}[th!b]
\centering
\caption{Two-Sample Kolmogorov-Smirnov Tests$^a$}
\label{tab: ks}.
\begin{tabular}{lrrrr}
\toprule
& \multicolumn{4}{c}{Days-since-reported (see Fig.~\ref{fig: reported})} \\
\cmidrule{2-5}
& Linux & FreeBSD & NetBSD & OpenBSD \\
\cmidrule{2-5}
Linux & & 0.566 & 0.530 & 0.315 \\
FreeBSD &&& 0.387 & 0.556 \\
NetBSD &&&& 0.481 \\
\cmidrule{2-5}
\\
& \multicolumn{4}{c}{Time-to-fix (see Fig.~\ref{fig: ttf})} \\
\cmidrule{2-5}
& Linux & FreeBSD & NetBSD & OpenBSD \\
\cmidrule{2-5}
Linux & & 0.410 & 0.536 & 0.404 \\
FreeBSD &&& \underline{0.368} & \underline{0.168} \\
NetBSD &&&& \underline{0.250} \\
\bottomrule
\multicolumn{5}{l}{\scriptsize $^a$~The values shown refer to $D$-statistics; underlined values denote $p \geq 0.05$.}
\end{tabular}
\end{table}

Quite a few fuzzing-generated crashes are still unresolved for all four kernels. Some of these are also relatively old: the medians for the pending crashes are 136 days for Linux, 51 days for FreeBSD, 53 days for NetBSD, and 119 days for OpenBSD. If a year is used as a threshold for long-lived bugs~\cite{Saha14}, there are as many as 80 long-lived bugs (crashes) in the Linux kernel but none in the three BSD kernels. Provided that some of these assumably have security implications, the results are enough to remark that implicit risks are present for running Linux kernels---though, explicit risks remain difficult to state; even when security consequences have been verified to be present, potential for exploitation is a different question. 

When turning to the TTF values, however, it seems that many of the already fixed bugs have been fixed relatively posthaste. The median TTF is 38 days for the Linux kernel and below 20 for all BSD kernels. A~plausible but hypothetical explanation is that crashes that are straightforward to interpret and debug yield also fast bug fixes, whereas the still open crashes may refer to particularly complicated ``heisenbugs''. The shapes of the empirical distributions for the TTF values are also visually similar for all four kernels, although the much higher amount of bugs for Linux distorts the image due to a few bugs that have taken a long time to fix. The unequal sample sizes warrant also some caution for interpreting the Kolmogorov-Smirnov test results in Table~\ref{tab: ks}. Nevertheless, the null hypotheses about equal distributions remain in force (\text{$p \geq 0.5$}) for the BSD kernels with respect to TTF but not DSR. Although a closer examination can be left for further work, a potential explanation may relate to the large amount of code shared between FreeBSD, NetBSD,~and~OpenBSD.

\subsection{Bug Types (RQ.2)}

The summary fields outputted by \textit{syzbot} can be used to group the types of bugs found from the Linux kernel into nine groups. The relative share of each type is shown in Table~\ref{tab: types}. Debug checks (WARNING) constitute the most common type. Also assertions (BUG) are quite frequently triggered together with use-after-free issues. The relatively large amount of general protection faults is explained by the use of x86-based Qemu virtual machines for the actual fuzzing. Interestingly, there exists also some variance between the open crashes and the already closed bugs in terms of the nine bug types listed.

\begin{table}[th!b]
\centering
\caption{Bug Types in Linux (\% Across Bug Types)}
\label{tab: types}
\begin{tabular}{lrrr}
\toprule
& Open & Closed & Both \\ 
\cmidrule{2-4}
WARNING & 18.9 & 18.9 & 18.9 \\ 
Use-after-free & 14.6 & 18.2 & 16.9 \\ 
General protection fault & 13.4 & 15.4 & 14.7 \\ 
BUG & 12.6 & 12.9 & 12.8 \\ 
Out-of-bounds read/write & 11.4 & 15.4 & 14.1 \\ 
Deadlock/stall & 3.4 & 2.9 & 3.1 \\ 
Uninitialized value & 5.2 & 6.2 & 5.8 \\ 
NULL pointer dereference & 1.3 & 0.8 & 1.0 \\ 
Other & 19.2 & 9.3 & 12.7 \\ 
\cmidrule{2-4}
$\sum$ & 100.0 & 100.0 & 100.0 \\
\bottomrule
\end{tabular}
\end{table}

\begin{table}[th!b]
\centering
\caption{Reviewed and Tested Commits (\% Within Bug Types)}
\label{tab: reviews}
\begin{tabular}{lrrr}
\toprule
& Reviewed & Tested & Both \\ 
\cmidrule{2-4}
WARNING & 11.4 & 9.8 & 21.3 \\ 
Use-after-free & 17.2 & 12.7 & 29.9 \\ 
General protection fault & 12.6 & 8.2 & 20.8 \\ 
BUG & 20.2 & 9.2 & 29.5 \\ 
Out-of-bounds read/write & 9.6 & 9.1 & 18.8 \\ 
Deadlock/stall & 5.1 & 2.6 & 7.7 \\ 
Uninitialized value & 4.8 & 4.8 & 9.6 \\ 
NULL pointer dereference & 27.3 & 0.0 & 27.3 \\ 
Other & 20.8 & 9.6 & 30.4 \\ 
\bottomrule
\end{tabular}
\end{table}

About 23\% of the commits for the fixed bugs have been either reviewed or tested by other developers. This rate seems sensible; a much higher share is presumably present for commits that introduce new features or make large restructurings. Some of the bugs behind the crashes may be simple enough to fix without code reviews. However, there exists some variance across the nine bug types. This observation can be seen from Table~\ref{tab: reviews}, which shows the share of reviewed and tested commits within each bug category. While fixing uninitialized variables may not necessitate reviews or testing, the low amount of reviews for deadlocks and stalls is a little surprising.

\subsection{Regression Analysis (RQ.3)}

The bug fixing times for the crashes reported by \textit{syzbot} for the Linux kernel tend to vary across the nine bug types for the open crashes (DSR) but not for the already fixed bugs (TTF). This observation can be seen from Table~\ref{tab: anova}. While the null hypotheses about equal means are rejected for both with plain ANOVA, the Leneve's test indicates heteroskedasticity across the bug types; therefore, the Welch's ANOVA shown in the third column provides a better estimator~\cite{Welch47}. Due to the unequal sample sizes and different operationalization, comparing the two cases does not make much sense, however. The observation regarding TTF nevertheless provides a good motivation to proceed into the more formal regression analysis.

\begin{table}[th!b]
\centering
\caption{Bug Fixing Times and Bug Types in Linux ($p$-values)}
\label{tab: anova}
\begin{tabular}{lcccc}
\toprule
& $n$ & ANOVA & Levene's test & Welch's ANOVA \\
\cmidrule{2-5}
DSR & \phantom{~}708 & $<$~0.001 & $<$~0.001 & $<$~0.001 \\
TTF & 1344 & $<$~0.001 & \phantom{$<~$}0.021 & \phantom{$<~$}0.084 \\
\bottomrule
\end{tabular}
\end{table}

The Cox's regression relies on the proportional hazards assumption. A violation of the assumption typically shows itself visually when two or more (Kaplan-Meier) survival curves drift into opposite directions, cross each other, or something alike. A formal test is also available, and problems are indicated by the test implemented in the R's \texttt{cox.zph  } function. To adjust for the problems, the deadlock/stall class was merged with general protection faults (cf.~Table~\ref{tab: types}), and the variables for reviews and tests (cf.~Table~\ref{tab: reviews}) were unified into a single variable. With this merging, the conventional $p < 0.05$ threshold is not crossed for the TTF case of interest. As can be further noted from Table~\ref{tab: prophaz}, the unified ``faults and deadlocks'' variable still indicates some potential problems, but the TTF model fitted seems sufficient for~interpretation. 

Thus, according to the final results in Table~\ref{tab: prophaz}, none of the coefficients for the reclassified bug types are statistically significant. All of the coefficients are also small in magnitude. This observation conforms with the ANOVA results. The only statistically significant coefficient is the one for the lines of code modified. The sign of this coefficient is also as expected, although also its magnitude is small. It can be also remarked that the dataset contains one large commit, but the results do not change much by excluding it.\footnote{~Identified by \texttt{ab8085c130edd65be0d95cc95c28b51c4c6faf9d}.} Although there may well be some potential interactions between the variables, it seems reasonable to tentatively conclude that only code churn seems to provide a signal about potential explanations for the time-to-fix values in the Linux kernel---yet even this signal is weak.

\begin{table}[th!b]
\centering
\caption{Proportional Hazards Assumption ($p$-values)}
\label{tab: prophaz}
\begin{tabular}{lcrr}
\toprule
&& DSR & TTF \\
\cmidrule{3-4}
Other (ref.) && -- & -- \\
WARNING && 0.147 & 0.364 \\
Use-after-free && 0.434 & 0.847 \\
Faults and deadlocks && 0.001 & 0.038 \\
BUG && 0.001 & 0.396 \\
Out-of-bounds read/write && 0.180 & 0.125 \\
Uninitialized value && 0.722 & 0.240 \\
NULL pointer dereference && 0.162 & 0.727 \\
\cmidrule{3-4}
Files modified && -- & 0.215 \\
Lines added && -- & 0.371 \\
Lines deleted && -- & 0.146 \\
\cmidrule{3-4}
Neither reviewed nor tested (ref.~) && -- & -- \\
Reviewed, tested, or both && -- & 0.829 \\
\cmidrule{1-4}
Global test && 0.001 & 0.052 \\
\bottomrule
\end{tabular}
\end{table}

\begin{table}[th!b]
\centering
\caption{Cox's Regression Results (TTF)}
\label{tab: cox}
\begin{tabular}{lcrr}
\toprule
&& Coef. & $p$-value \\
\cmidrule{2-4}
Other (ref.) && -- & -- \\
WARNING && -0.091 & 0.406 \\
Use-after-free && -0.073 & 0.508 \\
Faults and deadlocks && -0.058 & 0.599 \\
BUG && 0.040 & 0.737 \\
Out-of-bounds read/write && 0.088 & 0.444 \\
Uninitialized value && 0.227 & 0.112 \\
NULL pointer dereference && -0.002 & 0.994 \\
\cmidrule{3-4}
Files modified && 0.040 & 0.042 \\
Lines added && -0.001 & 0.313 \\
Lines deleted && -0.001 & 0.600 \\
\cmidrule{3-4}
Neither reviewed nor tested (ref.~) && -- & -- \\
Reviewed, tested, or both && -0.127 & 0.064 \\
\bottomrule
\end{tabular}
\end{table}

\section{Discussion}\label{sec: discussion}

\subsection{Conclusions}

This short exploratory paper examined three questions. The answers to the questions can be summarized as follows:

\begin{itemize}
\itemsep 2pt
\item{The answer to RQ.1 is two-fold: on one hand, many bugs particularly in the BSD kernels are fixed relatively quickly; on the other hand, there are over 800 open crashes for the four kernels examined, and many of these have been reported relatively long ago. In terms of the already fixed bugs, the fixing times are also similar between the BSD kernels. Much more crashes and bugs have reported for the Linux kernel, which partially also explains the observation that some of these are quite old.}
\item{Regarding RQ.2, the bug types are typical: assertions and debug checks (BUG and WARNING), use-after-frees (CWE-416), and general protection faults account for the majority of bugs in the sample. About 23\% of the fixed bugs in the Linux kernel have went through code review and/or testing. This low amount is a small surprise.}
\item{The answer to RQ.3 is clear: the bug types, code reviews, and additional testing do not seem to provide a solid statistical explanation for the bug fixing times in the Linux kernel. In fact, only one metric for code churn (files modified) seems to provide a weak statistical signal.}
\end{itemize}

Particularly the answer to RQ.3 is interesting: simple explanations for the bug fixing times do not seem plausible. This result provides a motivation for further empirical research.

\subsection{Limitations}

Four limitations can be briefly noted. The first is conceptual: the crashes reported by \textit{syzbot} may or may not equate to unique bugs. In other words, a single bug may be responsible for multiple crashes~\cite{Schumilo17, Ghafoor16}. Therefore, comparisons like the one in Fig.~\ref{fig: reported} should be interpreted with care. More generally, false positives and reproducibility are typical issues for automated continuous fuzzing~\cite{Carabas17}. The second limitation is empirical: bugs found via fuzzing should be compared to ``conventional'' bugs found and reported by humans as well as other types of automated kernel testing. To patch this limitation, comparisons such as the one in Table~\ref{tab: reviews} might be augmented with a random sample of other commits, for instance. The third limitation is longitudinal: only a snapshot limited to an early period of continuous kernel fuzzing was analyzed. The last limitation is methodological: inference with statistical significance should be approached with caution.

\subsection{Further Work}

Fuzzing has been extensively studied ever since the 1990s. In recent years kernel fuzzing has received particular attention. While also the potential benefits from machine learning have been considered, the considerations have concentrated on technical aspects, such as coverage, seed selection, and input generation~\cite{Saavedra19}. Very little work has been done on the practical software engineering side. This side includes questions about prioritization, cost-benefit analysis~\cite{Takanen08}, maintenance, and vulnerability tracking. Addressing these questions is not about strict verification, but more about providing heuristics for helping developers to analyze, fix, triage, and prioritize crashes reported by continuous fuzzing techniques. In this regard, three avenues for further empirical work seem worthwhile. 

First, the answer to RQ.3 allows to hypothesize that fixing bugs found by fuzzing may differ from more general bug fixing. If further research shows that there indeed is a difference, also practical improvements are easier to justify. Second, more research is required for providing hints to determine whether a crash has security implications. The crash dumps provide a good data source for this task. In addition to the general literature on mining of crash dumps, stack traces, and related outputs~\text{\cite{Ghafoor16, Schroter10}}, some recent work has also been done for automatically determining whether fuzzing-triggered crashes are exploitable based on the associated crash dumps~\cite{YanLu17}. Third, mining of commit messages may provide further insights about security implications. There is also some work for mapping textual information to CWEs~\cite{Ruohonen18TIR}, for instance. Further data mining may also help at explaining the bug fixing times; indeed, both stack traces~\cite{Schroter10} and (security) bug severity~\cite{Chinthanet19} have been observed to correlate with fixing~times.

\pagebreak
\clearpage
\balance
\bibliographystyle{IEEEtran}


\end{document}